\documentclass[%
 aip,
 amsmath,amssymb,
 reprint,%
]{revtex4-1}

\usepackage{graphicx}
\usepackage{dcolumn}
\usepackage{bm}

\usepackage[utf8]{inputenc}
\usepackage[T1]{fontenc}
\usepackage{mathptmx}
\begin{document}

\preprint{AIP/123-QED}

\title{Laser Intensity Noise Suppression for Preparing Audio-Frequency 795 nm Squeezed Vacuum State of Light at Rubidium D1 Line
}

\author{Lele Bai$^{1}$, Xin Wen$^{1,2}$, Yulin Yang$^{1}$, Jun He$^{1,3}$, Junmin Wang$^{1,3,*}$}
\affiliation{$^1$State Key Laboratory of Quantum Optics and Quantum Optics Devices, Institute of Opto-Electronics, Shanxi University, Taiyuan 030006, China\\
$^2$Department of Physics, Tsinghua University, Beijing 100084, China\\
$^3$Collaborative Innovation Center of Extreme Optics of the Ministry of Education and Shanxi Province, Shanxi University, Tai Yuan 030006, China\\
$^*$wwjjmm@sxu.edu.cn}

\date{\today}

\begin{abstract}
Laser intensity noise suppression has essential effects on preparation and characterization of the audio-frequency squeezed vacuum state of light based on a sub-threshold optical parametric oscillator (OPO).We have implemented two feedback loops by using relevant acousto-optical modulators (AOM) to stabilize the intensity of 795-nm near infrared (NIR) fundamental laser and 397.5-nm ultraviolet (UV) laser generated by cavity-enhanced frequency doubling.Typical peak-to-peak laser intensity fluctuation with a bandwidth of $\sim10$ kHz in a half hour has been improved from $\pm7.45$$\%$ to $\pm0.06$$\%$ for 795-nm NIR laser beam, and from $\pm9.04$$\%$ to $\pm0.05$$\%$ for 397.5-nm UV laser beam, respectively. The squeezing level of the squeezed vacuum state at 795 nm prepared by the sub-threshold OPO with a PPKTP crystal has been improved from -3.3 to -4.0 dB around 3$\sim$9 kHz of audio analysis frequency range.
\end{abstract}

\keywords{laser intensity noise; feedback loop; squeezed vacuum state of light; acousto-optical modulator (AOM); audio frequency}

\maketitle

\section{\label{sec:level1} Introduction}
The first squeezed vacuum state of light based on an optical parametric oscillator (OPO) was realized by Wu et al.\cite{Note1}, since its noise can be lower than the shot noise level (SNL) in a certain quadrature component. The squeezed vacuum state of light has great significance in gravitational wave detection \cite{Note2, Note3, Note4} and can be used in quantum communication, quantum storage, and precision measurement \cite{Note5, Note6, Note7, Note8}.At the analysis frequency of MHz, the influence of laser intensity noise upon producing and detecting the squeezed vacuum state of light can be neglected as its background noise is pretty quiet and the SNL can be achieved easily \cite{Note9, Note10, Note11, Note12, Note13}. The squeezed vacuum state of light at audio frequency \cite{Note14} plays a vital role in certain areas of research, such as gravitational wave detection \cite{Note2, Note3, Note4}, magnetic measurements \cite{Note15, Note16, Note17, Note18, Note19}, and biological measurement \cite{Note20}, whose sensitivity is limited by the squeezing level due to the coupling of low frequency noise. With the further exploration of audio-frequency squeezed vacuum state of light, it will open a new door for more fields.

Traditional methods of suppressing the laser intensity noise are injection locking \cite{Note21}, opto-electronical feedback \cite{Note22}, and mode cleaner \cite{Note23}. Some new schemes have also been proposed, such as based on the second-harmonic component of the photo-elastic modulator detection \cite{Note24} and controlling the temperature and mechanical vibration of electro-optical crystal to realize the long-term and wide-band laser intensity stabilization \cite{Note25}. Recently, a nonclassical way to improve the achievable power stability by the injection of squeezed vacuum field of light has been demonstrated \cite{Note26}. All the above-mentioned schemes are used for various applications which cover the laser writing system \cite{Note27}, coherent population trapping clock \cite{Note28}, and other optical devices \cite{Note29}. Here, we aim to reduce the laser intensity noise to weaken its negative impact on preparation and detection of squeezed light at audio frequency range \cite{Note30, Note31}.

For preparation and improvement of audio-frequency 795-nm squeezed vacuum state of light based on a sub-threshold OPO, we designed two sets of feedback loop systems with acousto-optical modulators (AOM) to obtain high-stability 795-nm near infrared (NIR) and 397.5-nm ultra-violet (UV) lasers. Actually, the system will not only play a vital role in the preparation and detection of audio-frequency squeezed vacuum state of light, but also be beneficial to other related optical experiments. In principle, as the laser intensity noise of 795-nm NIR laser is suppressed to SNL, the effect of classical noise on the squeezed vacuum state of light can be eliminated accordingly. However, in practice, a series of nonlinear transformations in the process of frequency doubling will bring extra noise to the second-harmonic wave, no matter how stable the fundamental wave is. This is why the laser fluctuation of 397.5-nm UV laser generated by the second-harmonic generation (SHG) is normally greater than that of 795 nm and we have to use another system to suppress its intensity noise at the same time. Finally, we realized different suppression effects of laser intensity in the case of different feedback bandwidths and different sampling powers. The feedback effects are characterized by a digital multimeter in the time domain and a fast-Fourier-transformation (FFT) dynamic spectrum analyzer in the frequency domain. With feedback control, typical peak-to-peak laser intensity fluctuation is less than 1/100 of the original one in 30 min. With the help of the NIR and UV laser's intensity noise suppression, squeezing level of the squeezed vacuum state of 795-nm light prepared by OPO with a PPKTP crystal is improved from -3.3 to -4.0 dB around 3$\sim$9 kHz audio frequency.

\section{Demand for Laser Intensity Noise Suppression for Preparing Audio-Frequency Squeezed Vacuum State of Light}

For the singly-resonant OPO (no seed light) or the singly-resonant optical parametric amplifier (OPA) (with the seed light), the corresponding amplitude (+) and phase (-) quadrature variance $V^{\text{$\pm$}}_{sqz}$($\omega$) can be expressed as follows \cite{Note32}:
\begin{eqnarray}
V^{\text{$\pm$}}_{sqz}(\omega)&=& \frac{{C_{s}V^{\text{$\pm$}}_{s}(\omega)}+{C_{l}V^{\text{$\pm$}}_{l}(\omega)}+{C^{\text{$\pm$}}_{v}V^{\text{$\pm$}}_{v}(\omega)}}{\vert D^{\text{$\pm$}}(\omega) \vert ^2} \nonumber \\
& &+\frac{{\alpha}^{2}[{C_{p}V^{\text{$\pm$}}_{p}(\omega)}+{C^{\text{$\pm$}}_{\text{$\Delta$}}V_{\text{$\Delta$}}(\omega)}]}{\vert D^{\text{$\pm$}}(\omega) \vert ^2},
\end{eqnarray}
where$V^{\text{$\pm$}}_{s}(\omega)$ is the noise of seed fields, $V^{\text{$\pm$}}_{l}(\omega)$ is the vacuum fluctuation caused by intra-cavity loss, $V^{\text{$\pm$}}_{v}(\omega)$ is the vacuum fluctuation from output coupler, $V^{\text{$\pm$}}_{p}(\omega)$ is the noise from pump fields, and $V_{\text{$\Delta$}}(\omega)$ is the noise of detuning fluctuation in the cavity, respectively. Additionally, $C_{s}$, $C_{l}$, $C^{\text{$\pm$}}_{v}$, $C_{p}$ and $C^{\text{$\pm$}}_{\text{$\Delta$}}$ are their respective coefficients; $D^{\text{$\pm$}}(\omega)$ is the description of the performance of the OPO cavity, independent of the light field; $\alpha^2$ is the number of photons of the seed light into the OPO.

In theory, the OPO mode ($\alpha^2=0$) can avoid the destruction of the audio-frequency squeezed vacuum state as it eliminates the possibility that the obvious fluctuation of pump fields and the noise of detuning fluctuation are brought into the audio-frequency squeezed vacuum state of light. In order to reduce the negative impact on the audio-frequency squeezed state of light, the polarization of lock beam with a certain frequency shift (140 MHz) should be P polarization, perpendicular to the pump light (S polarization), and the propagation direction should be opposite to the pump light. If the scattering and residual reflection of the lock beam on the surface of nonlinear crystal (PPKTP crystal in our experiment) inside OPO cavity are not considered, OPO runs as an ideal OPO mode for generation of audio-frequency squeezed vacuum state of light. However, actually, the lock beam's polarization degree is limited and that is normally not a pure P-polarization beam as we hope, and it contains a residual weak S-polarization component. Due to roughness of surface of nonlinear crystal (PPKTP) and imperfect anti-reflection coating, the scattering and residual reflection of the lock beam cannot be simply ignored. Some fundamental-wave photons with S polarization along the direction of OPO's pump beam actually serve as "seeding signal" ($\alpha^2\ne0$), and will partially blur the audio-frequency squeezed state of light. This point was addressed by McKenzie et al. \cite{Note31}], and if the seeding signal's power is >1 nW, the audio-frequency squeezing level will be degraded, and OPO actually will run as an OPA mode. Since intensity noise of "seeding signal" is the key factor affecting squeezing level, introducing the feedback system is particularly important, which not only can reduce classical noise but also can benefit the stable locking of OPO/OPA cavity. In addition, the intensity of pump beam has close relationships with the audio-frequency squeezed state of light, which not only brings considerable classical gain, but also introduces some extra noise including thermal instability and NIR laser's absorption induced by UV laser, especially when the intensity noise is large \cite{Note8}. The drift of thermal noise and other classical noise can be mitigated by stabilizing its intensity fluctuation, which is also more conductive to the stable output of squeezed light for a long time.

Yang et al. \cite{Note11} have analyzed the intensity noise of local oscillator (LO) beam and the common mode rejection ratio (CMRR) of balanced homodyne detector (BHD) at different analysis frequency. Similarly, in our detection system, as the intensity noise of LO is higher than the SNL, and the CMRR of BHD is limited ($\sim35$ dB), the output squeezing level of OPO/OPA cannot be detected accurately. There are two ways to improve the accuracy of the squeezing level, one is to increase BHD's CMRR to cancel out the relative intensity noise of LO higher than the SNL in the audio frequency range, which is difficult to achieve due to the technical and electronic circuit barriers; the other is to suppress the relative intensity noise of LO, which makes it possible for BHD to offset the common-mode noise and easier for us to implement.

\section{Experimental Setup of Feedback System for Laser Intensity Noise Suppression}

In the experiment of preparing and detecting audio-frequency squeezed vacuum state of 795-nm light, the laser source we used is a continuous-wave single-frequency tunable Ti:Sapphere laser (Model SolsTiS, M Squared, UK), which can be tuned to 795-nm NIR and the output power is about 2.5 W. As shown in Figure 1(a), the output beam is divided into two parts, one part needs to be stabilized by the 795-nm NIR laser feedback system and then used as the lock beam of OPO and the LO beam. The other part is used to realize 397.5-nm UV laser for pumping OPO via cavity-enhanced SHG cavity with a LBO bulk crystal while two high-reflectivity mirrors (F) serve as switch of the feedback system.

The feedback device consists of two parts, in-loop and out-loop parts, which are respectively used for servo control and monitor, shown in Figure 1(b). The in-loop part of the device mainly includes AOM, polarization-maintaining fiber (PMF), polarization beam splitter (PBS) cube, beam splitter (BS) plate, sampling photodetector (PD1), and servo control, for details see reference \cite{Note33}. According to the Bragg diffraction mode of AOM, there is only a first-order diffraction beam and a zero-order beam after passing through AOM and its temperature is controlled at room temperature accurately by an oven to prevent thermal instability produced by the interaction between crystal and laser from increasing the intensity fluctuation of laser. The first-order diffraction beam is blocked with dump and the zero-order diffraction beam with no frequency shift is injected into a short PMF fixed by ultra-stable mounts, which not only can act as a filter to clean the spatial mode, but also reduce the pointing instability of the beam. In order to reduce the fluctuation of polarization, a PBS with an extinction ratio of 3000:1 was used. The laser reflected by BS1 is utilized as in-loop laser sampling, and the transmitted laser is used for monitor and later use. The sampling laser with intensity fluctuation is transformed into an electrical signal via PD1 and a servo control to produce a signal to drive AOM. The PD2 is used to characterize the noise of laser intensity.

\begin{figure}[!htbp]
\begin{center}
\includegraphics[height=45mm,width=85mm]{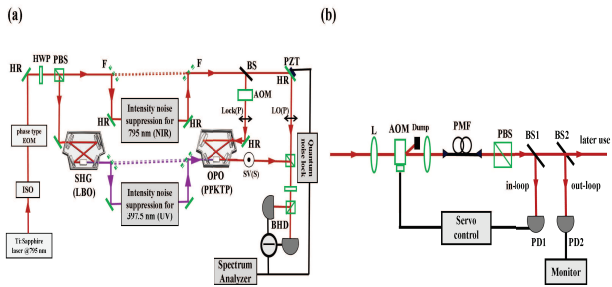}
\caption{\label{fig:epsart} (a) Schematic diagram of experimental setup for preparing the squeezed vacuum state of 795-nm light at audio frequency. The dotted line path means without feedback and the solid line represents noise suppression via feedback. (b) Diagram of experimental setup of the AOM feedback system. Both 795-nm NIR and 397.5-nm UV laser's AOM feedback systems are similar. ISO: Optical isolator; EOM: Electro-optical modulator; NIR: Near infrared; UV: Ultraviolet; HWP: Half-wave plate; PBS: Polarization beam splitter cube; BS: Beam splitter plate; L: Lens; AOM: Acousto-optical modulator; D: Dump; PMF: Polarization-maintaining fiber; PD: Photo-detector; HR: High-reflectivity mirror; SHG: Second-harmonic generation cavity; OPO: Optical parametric oscillator cavity; SV: Squeezed vacuum state of light; LO: Local oscillator beam; S: S polarization; P: P polarization; BHD: Balanced homodyne detectors. }
\end{center}
\end{figure}

\section{Experimental Results and Discussion}
\subsection{Suppression of 795-nm NIR and 397.5-nm UV Laser Intensity Noise}

The radio-frequency power (omitted in Figure 1) driving AOM is controlled by a servo system according the adjustment of Bragg diffraction of AOM to laser intensity which could be thus stabilized. Finally, laser power can be stabilized to a different level with disparate feedback bandwidth. The following two sets of feedback results for 795-nm NIR laser are obtained by changing the bandwidth of PD1 and adjusting parameters of the servo system, as shown in Figure 2(a)-(d). For (a) and (b), the bandwidth of PD1 is set to DC-300 kHz, for (c) and (d), the bandwidth of PD1 is set to DC-100 kHz. The corresponding bandwidth of PD2 is DC-300 kHz, which can cover a wide range of frequencies that we care about. The optical power reached PD1 and PD2 are $\sim2$ and $\sim5$ mW, respectively. For Figure 2, the feedback bandwidth is mainly determined by the parameter of integral time in the servo control and selection of PD1's bandwidth. In addition, the response frequency of AOM and attenuator also plays an auxiliary role in the feedback bandwidth. Here, the effect of the AOM (Model 3080-122 @ 795 nm, Model 1-M110 @ 397.5 nm, Cooch $\&$ Housego, Somerset, UK) and attenuator (Model ZAS-3+, Mini-circuits, Brooklyn, USA) on the feedback bandwidth can be ignored as their response frequency is higher than MHz. We adopt a digital multimeter (Model DM3068, RIGOL, China) to record the output of PD2 for 30 min in time domain. To characterize relative laser intensity noise in frequency domain a FFT dynamic spectrum analyzer (Model SR785, SRS, USA) is employed.

\begin{figure}[!htbp]
\begin{center}
\includegraphics[height=45mm,width=65mm]{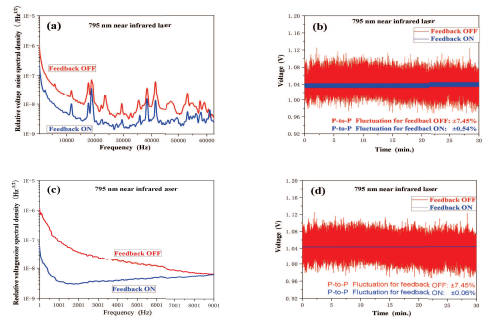}
\caption{\label{fig:epsart}Comparison of laser intensity noise of the 795-nm NIR laser for feedback system on and off with different feedback bandwidths. The effective feedback bandwidth is $\sim60.0$ kHz for (a) and (b), and $\sim9.0$ kHz for (c) and (d), respectively. In time domain, typical peak-to-peak fluctuation of 795-nm NIR laser intensity is improved from $\pm7.45$$\%$ to $\pm0.54$$\%$ (b) and $\pm0.06$$\%$ (d), respectively. The 795-nm NIR laser intensity's stability is improved by a factor of $\sim13$ (b) and $\sim120$ (d), respectively.}
\end{center}
\end{figure}

Analyzing the above-mentioned experimental results, the narrower feedback bandwidth is, the more obvious the suppression effect will be. However, not all bandwidth can be suppressed due to the limitation of feedback loop. In addition, in order to map the laser intensity noise on the feedback detector to be more real, we try to let as much optical power as possible to reach the sampling detector to make feedback better, but it must be below the saturation level of the detector. Responses at five different sampling optical powers are shown in Table 1.

\begin{table*}
\caption{\label{tab:table1}Residual peak-to-peak fluctuation of 795-nm NIR laser intensity with feedback control vs. optical power of the sampling beam.}
\begin{ruledtabular}
\begin{tabular}{cccccc}
Power of the sampling beam (mW)& $\sim0.5$ & $\sim0.8$ & $\sim1.0$ & $\sim1.6$ & $\sim2.0$\\
\hline
Residual peak to-peak fluctuation of 795-nm \\
NIR laser intensity with feedback control & $\pm0.32$$\%$ & $\pm0.24$$\%$ & $\pm0.17$$\%$ & $\pm0.11$$\%$ & $\pm0.06$$\%$\\
\end{tabular}
\end{ruledtabular}
\end{table*}

According to the same principle, we set up a feedback system for 397.5-nm UV laser beam, as shown in Figure 1(b). We changed the type of AOM and in-loop parameter setting. The results are shown in Figure 3(a),(b), and the corresponding 397.5-nm UV sampling power is $\sim5.0$ mW, the bandwidth and gain factor of PD1 are DC $\sim100$ kHz, and 3, respectively.

\begin{figure}[!htbp]
\begin{center}
\includegraphics[width=80mm]{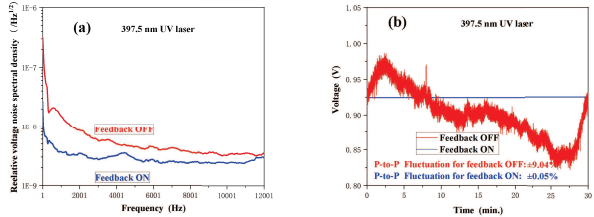}
\caption{\label{fig:epsart}Comparison of laser intensity noise of 397.5-nm UV laser beam for feedback system on and off. (a) The effective feedback bandwidth is $\sim11.5$ kHz. (b) In time domain, typical peak-to-peak fluctuation of 397.5-nm UV laser intensity is improved from $\pm9.04$$\%$ to $\pm0.05$$\%$. The stability has been improved by a factor of $\sim180$.}
\end{center}
\end{figure}

\subsection{Preparing Audio-Frequency Squeezed Vacuum State of 795-nm Light}

The experimental setup to prepare audio-frequency squeezed vacuum state of 795-nm light has four parts, 795-nm single-frequency laser system, SHG, OPO, and BHD, as shown in Figure 1(a). On the basis of ~1 W of fundamental optical power, the second-harmonic at 397.5-nm of $\sim300$ mW, can be generated, with a typical doubling efficiency of $\sim30$$\%$. After shaping the spot and stablizing laser intensity, the 397.5-nm UV laser power is $\sim60$ mW,which is below the pump threshold of OPO and it can provide the classical gain of $\sim8$, in which case OPO can provide stable output of squeezed vacuum state of light. At the same time, after optimizing the intensity and beam quality of LO beam, the squeezed vacuum state of light can be detected by the BHD which has 99.8$\%$ of interference visibility. The results of squeezed vacuum state of light in range of 2 $\sim15$ kHz have been achieved with the quantum noise locking technique, as shown in Figure 4. Comparing the level of squeezing and anti-squeezing in two cases of feedback on and off indicates the feedback bandwidth is 9.3 kHz. In feedback bandwidth, the squeezing level is improved by 0.7 dB after the suppression of laser intensity noise. Unfortunately, due to the complex noise introduced by the circuit in loop of system, additional noise appears at some certain frequencies, shown as several peaks. The squeezing level is not improved for the analysis frequency beyond the feedback bandwidth. On the contrary, the anti-squeezing level is correspondingly increased by 0.7 dB.

\begin{figure}[!htbp]
\begin{center}
\includegraphics[width=80mm]{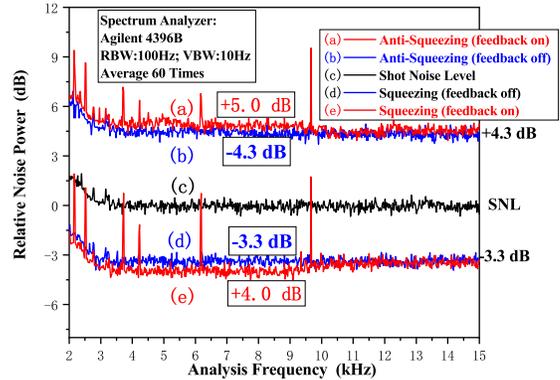}
\caption{\label{fig:epsart}Relative noise power spectra normalized to the shot noise level (SNL) for the squeezing phase and the anti-squeezing phase. The two cases with and without both the 795-nm NIR and the 397.5-nm UV intensity noise feedback loops are compared. The traces (a) and (e) indicate the anti-squeezing of +5.0 dB and the squeezing of -4.0 dB for the case with feedback. The traces (b) and (d) indicate the anti-squeezing of +4.3 dB and the squeezing of -3.3 dB for the case without feedback. The trace (c) shows SNL. Clearly in the analysis frequency range from $\sim3.3$ to $\sim9.3$ kHz the squeezing is improved from -3.3 to -4.0 dB with help of both the 795-nm NIR and the 397.5-nm UV intensity noise feedback loops. In fact, there is a significant loss in the path from the OPO output coupling mirror to the BHD, including the round trip losses ($\sim3$$\%$) of light and limited quantum efficiency ($\sim97$$\%$) of BHD we utilized, which directly lead to an underestimate of the squeezing level. Therefore, we can infer that the true squeezing level of audio-frequency squeezed vacuum state of light is 0.2-0.3 dB higher than we detected. The noise power spectra are measured with a spectrum analyzer (Aglient 4396B, USA) and averaged for 60 times. The resolution bandwidth (RBW) is 100 Hz, and video bandwidth (VBW) is 10 Hz.}
\end{center}
\end{figure}

\section{Conclusion}
We used the zero-order diffraction light of AOM instead of the first-order diffraction for sampling feedback, and successfully suppressed the 795-nm NIR and 397.5-nm UV lasers' intensity fluctuation involved in the preparation of squeezed vacuum state of light. After stabilizing, typical peak-to-peak laser intensity fluctuation is less than 1/100 of the original one. The relationship between the bandwidth, sampling power, and feedback effect is investigated. The narrower the bandwidth is, the more obvious the suppression effect is; and the larger the sampling power is, the better the feedback effect will be on the premise that the optical power reaching the detector is not saturated. By effectively suppressing intensity noise, not only the squeezing level can be improved in the feedback bandwidth, but also the stable continuous output of the squeezed vacuum state of light can be guaranteed. It is expected to achieve higher squeezing level by reducing the loss of OPO cavity and improving the detection system, which has great application value in precise measurement. In addition, further stabilizing the laser intensity and phase to close SNL and improving the detector's performance will yield the squeezed vacuum state of light at much lower analysis frequency.

\section*{Acknowledgments}
This research is partially funded by The National Key R$\&$D Program of China, grant number 2017YFA0304502, the National Natural Science Foundation of China, grant number 11974226, 61905133, 11774210, and 61875111, and the Shanxi Provincial 1331 Projects for Key Subjects Construction.

\nocite{*}
\end{document}